%% file: paper.tex
\documentclass[10pt,letterpaper]{article}
\usepackage{fullpage}

\input{header}

\title{Evaluating the Generalization Capabilities of \\ Large Language Models on Code Reasoning}

\author{
  \parbox{0.45\textwidth}{\centering Rem Yang\\MIT CSAIL\\\texttt{remyang@csail.mit.edu}}
  \parbox{0.45\textwidth}{\centering Julian Dai\\Brown University\\\texttt{julian\_dai@brown.edu}}
  \\[5ex]
  \parbox{0.45\textwidth}{\centering Nikos Vasilakis\\Brown University\\\texttt{nikos@vasilak.is}}
  \parbox{0.45\textwidth}{\centering Martin Rinard\\MIT CSAIL\\\texttt{rinard@csail.mit.edu}}
}

\date{}

\begin{document}
\maketitle

\begin{abstract}
    \input{abstract}
\end{abstract}

\input{introduction}
\input{related}
\input{dsl_mutation}
\input{benchmarks}
\input{experiments}
\input{results}
\input{conclusion}

\bibliographystyle{plainnat}
\bibliography{paper}
\clearpage

\appendix
\input{appendix}

\end{document}

%% file: header.tex
\usepackage{algorithm}
\usepackage{algpseudocode}
\usepackage{amsmath, amsthm, amssymb}
\usepackage{bm}
\usepackage{booktabs}
\usepackage{caption}
\usepackage{enumitem}
\usepackage{float}
\usepackage{listings}
\usepackage{makecell}
\usepackage{multicol}
\usepackage{multirow}
\usepackage{pgfplots}
\usepackage{pythonhighlight}
\usepackage{relsize}
\usepackage{subcaption}
\usepackage{tcolorbox}
\usepackage{wrapfig}
\usepackage{xcolor}

\definecolor{snsBlue}{HTML}{1f77b4}
\definecolor{snsOrange}{HTML}{ff7f0e}

\usepackage[sort]{natbib}
\setcitestyle{authoryear}
\usepackage[colorlinks=true, citecolor=snsBlue, linkcolor=snsBlue, urlcolor=snsBlue]{hyperref}

\tcbset{
    colback=gray!5,
    colframe=snsBlue
}
\pgfplotsset{compat=1.18}

%% file: abstract.tex
We assess how the code reasoning abilities of large language models (LLMs) generalize to different kinds of programs. We present techniques for obtaining in- and out-of-distribution programs with different characteristics: code sampled from a domain-specific language, code automatically generated by an LLM, code collected from competitive programming contests, and mutated versions of these programs. We also present an experimental methodology for evaluating LLM generalization by comparing their performance on these programs. We perform an extensive evaluation across 10 state-of-the-art models from the past year, obtaining insights into their generalization capabilities over time and across different classes of programs. Our results highlight that while earlier models exhibit behavior consistent with pattern matching, the latest models exhibit strong generalization abilities on code reasoning.

%% file: introduction.tex
\section{Introduction}
Large language models (LLMs) are increasingly used in software engineering and have demonstrated high performance on a variety of code-related benchmarks~\citep{austin2021program,chen2021evaluating,liu2023your,gong2024evaluation,jimenez2023swe,jain2025livecodebench}. However, the extent to which these models can effectively reason about program behavior remains an open question. Investigating LLM performance on code reasoning tasks~\citep{gu2024cruxeval,liu2024codemind,chen2025reasoning,jain2025livecodebench} is therefore of critical importance, especially since the ability to reason about programs has been shown to enhance the performance of LLMs on code generation~\citep{ni2023lever}, program repair~\citep{ni2024next}, and more general tasks that involve logical, mathematical, scientific, and commonsense reasoning~\citep{li2025codei}.

A key challenge in evaluating LLMs on coding tasks is distinguishing generalization from memorization and/or pattern matching across different classes of programs. Previous works~\citep{dong2024generalization,riddell2024quantifying} have shown that HumanEval~\citep{chen2021evaluating} and MBPP~\citep{austin2021program}, two of the most commonly used code generation benchmarks, contain a large overlap with LLM training data. On competitive programming benchmarks, LLMs exhibit a drop in performance on problems after their knowledge cutoff date, likely due to data contamination~\citep{huang2024competition,roberts2024to,jain2025livecodebench}. However, understanding the behavior of LLMs on different distributions of programs and how to effectively measure their generalization abilities has not yet been extensively studied for code reasoning.

To address this gap, we systematically evaluate LLM code reasoning performance on in- vs. out-of-distribution programs. We propose two general methods for obtaining new programs that are likely out-of-distribution: (1) sampling programs from a domain-specific language (DSL) with a combinatorially large space of programs and (2) applying mutation operators~\citep{papadakis2019mutation,jia2010analysis}. These mutation operators apply small syntactic changes that (approximately) preserve the complexity of a given program while changing its semantics. 

We create three datasets containing different classes of programs: (1) \textbf{LLM-List}: common list-processing functions (e.g., sort, search) generated by an LLM (which we expect to be in-distribution); (2) \textbf{DSL-List}: list-processing functions sampled from a DSL (which we expect to be out-of-distribution); and (3) \textbf{LeetCode}: human-submitted LeetCode contest solutions, which we separate into those before and after the knowledge cutoffs for the LLMs we evaluate. For each dataset, we further obtain mutated versions of its programs.

We perform two sets of experiments on these datasets:
\begin{enumerate}
    \item \textbf{Execution Prediction.} These experiments present an LLM with a program and a test input and instruct it to predict the output. We measure whether a model correctly predicts the outputs of the original and mutated programs. We also introduce the novel metric of \emph{reversion}, which measures whether a model predicts the output of the other version of its given program; e.g., reversion on a mutated program means that a model predicts the output of its corresponding original program.

    Substantially higher correctness on original programs than on mutated programs and high reversion on mutated programs are consistent with (1) original programs being in-distribution, (2) mutated programs being out-of-distribution, and (3) a model using pattern matching to predict program outputs. Comparable performance on original and mutated programs with little reversion is consistent with either both sets of programs coming from similar distributions or a model generalizing to out-of-distribution programs. To distinguish these cases, we design a second set of experiments.

    \item \textbf{Execution Choice.} These experiments present an LLM with both the original and mutated versions of a program and instruct it to (1) identify the version it is more confident in reasoning about and (2) predict the output for that program. We emphasize that we present the two programs without mentioning the existence of mutations. We measure LLM preference toward choosing the original or mutated program and its correctness and reversion when reasoning about the chosen program.

    The key insight is that a consistent preference for selecting the original programs is consistent with them being in-distribution and mutation moving them out-of-distribution; selecting the original and mutated programs at a similar rate indicates that the two versions' distributions are similar.
\end{enumerate}

\paragraph{Empirical Findings.} We perform an extensive evaluation on a suite of state-of-the-art open- and closed-access models released between November 2023 and March 2025, spanning the latest reasoning models --- QwQ~\citep{qwq}, DeepSeek-R1~\citep{guo2025deepseek}, and o3-mini~\citep{o3mini} --- and more traditional LLMs --- DeepSeek-Coder~\citep{guo2024deepseek}, Qwen2.5-Coder~\citep{qwen2.5}, GPT-4o-mini~\citep{gpt4omini}, and GPT-4o~\citep{hurst2024gpt}. We summarize our main empirical findings below.
\begin{enumerate}
    \item \textbf{How has LLM generalization capability on code reasoning changed over time?} Our results indicate a substantial increase in LLM generalization on code reasoning over the past year. Earlier LLMs from a little over a year ago exhibit poor correctness on all problems except original LLM-List programs; in stark contrast, the most recently released reasoning LLMs exhibit near-perfect correctness on all problems.

    \item \textbf{How does LLM performance vary across different classes of programs?} On LLM-List, original correctness is high for all models; mutated correctness is low and reversion is high for earlier models, while the latest models obtain near-perfect mutated correctness and near-zero reversion. On DSL-List, original and mutated correctness is comparable and reversion is low for each model; earlier models achieve poor performance while the latest models are near perfect. On LeetCode, earlier LLMs exhibit a noticeable gap in correctness between problems before and after cutoff, while there is essentially no gap for the latest reasoning LLMs; correctness on mutated problems compared to original problems is much lower (and reversion is high) for earlier models, while later models exhibit only a small drop in correctness with low reversion. We also further examine these results as a function of program complexity in our evaluation.

    These results are consistent with LLM-List programs being in-distribution, DSL-List programs being out-of-distribution, LeetCode programs being somewhat in-distribution, and mutation moving in-distribution programs out-of-distribution.
    They are also consistent with earlier models relying on pattern matching and the latest models generalizing to reason successfully for all of our benchmark programs. The results also underline that LLM code reasoning performance on substantially in-distribution problems is not representative of their generalization abilities, and that DSL-sampling and mutation are effective ways to obtain programs that better measure generalization capabilities. 

    Furthermore, the diminishing performance gap between before- and after-cutoff LeetCode problems as LLMs become more capable highlights that the technique of evaluating LLMs' code reasoning abilities on problems after cutoff may not be sufficient for creating out-of-distribution problems; our hypothesis is that human-written code, whether produced before or after knowledge cutoff, contains common programming patterns.

    \item \textbf{Can LLMs distinguish between original and mutated programs, and how does choosing to reason about the original or mutated programs affect their performance?} On LLM-List, all models prefer original programs to mutated programs and achieve high correctness and minimal reversion when reasoning about the original programs. For the (relatively few) cases where the models choose the mutated version, correctness is poor and reversion is substantial (with the exception of GPT-4o and the latest reasoning LLMs). On DSL-List, all models exhibit little preference for original versus mutated versions and achieve comparable correctness for both versions when selected. On LeetCode, all models prefer reasoning about the original programs. Correctness (on both selected original and mutated versions) generally improves as LLM release date increases; reversion when reasoning about the mutated version is high for most traditional models, while being nonexistent for reasoning LLMs.

    These results are consistent with LLM-List programs being in-distribution for all models (including the latest ones) and mutation moving in-distribution programs out-of-distribution (while not affecting the distribution of already out-of-distribution programs) for all models. They also further highlight the reliance of earlier models on pattern matching, while accentuating the generalization abilities of the latest reasoning LLMs on code reasoning.
\end{enumerate}

%% file: related.tex
\section{Related Work}
\subsection{Data Contamination and Memorization}
HumanEval~\citep{chen2021evaluating} and MBPP~\citep{austin2021program} are two of the most widely used evaluation benchmarks for code LLMs, designed to test their ability to synthesize relatively simple Python functions given a natural language description. However, these benchmarks have been found to suffer from severe data contamination issues, with a large portion of their solutions (or similar code) seen during training for many LLMs~\citep{dong2024generalization,riddell2024quantifying}. Furthermore, \citet{huang2024competition} and \citet{roberts2024to} show that while competitive programming questions are effective at evaluating LLMs, there is a large disparity between the performance of models before and after their knowledge cutoffs. These concerns have motivated the development of live benchmarks~\citep{jain2025livecodebench, white2025livebench, jimenez2023swe}. These benchmarks aim to bypass data contamination by continually updating their set of problems from a given data source, and by evaluating LLMs on problems released after their knowledge cutoffs. For instance, \citet{jain2025livecodebench} continually draws competitive programming questions from sources such as LeetCode and Codeforces contests. Another line of work reveals that LLMs memorize large amounts of code and are prone to pattern matching~\citep{yang2024unveiling, carlini2021extracting, rabin2023memorization}. These findings further highlight the importance of evaluating LLM performance on out-of-distribution scenarios that test their genuine reasoning abilities. Our work addresses data contamination issues by using DSL sampling and program mutation as mechanisms to create out-of-distribution problems with which to evaluate LLM generalization. Furthermore, contrary to prior work, our evaluation highlights that on the task of code reasoning, (1) evaluating LLMs on problems after their cutoff date may not be sufficient for testing their generalization abilities and (2) recent reasoning models possess surprisingly strong generalization abilities, providing evidence that LLMs are capable of going beyond memorizing or performing surface-level pattern matching on their training data.

\subsection{LLM Reasoning Capabilities in Other Domains}
With the rapid rise in LLM performance across many domains, recent research has studied how well the apparent reasoning abilities of LLMs generalize. \citet{mirzadeh2025gsm} study LLM reasoning on math problems when names and numerical values in the problems are altered and when irrelevant facts are introduced; they find that LLMs are brittle, with performance varying significantly in response to these perturbations. Similarly, \citet{jiang2024peek} show that LLMs' reasoning processes exhibit token bias: their performance on logical puzzles is affected by entities (e.g., names and numbers) not relevant to the underlying logic required to solve the problem. \citet{qi2025quantifying} investigate how an LLM's ability to solve algorithmic tasks differs between in-distribution and out-of-distribution input values; however, this work does not evaluate LLMs on code, and their tasks are instead phrased in natural language. These works start with a fixed underlying structure that solves a given problem, then evaluate the ability of LLMs to ignore irrelevant perturbations or augmentations to this structure. In contrast, the mutations that we apply are directly relevant to the reasoning task because they change the correct result. Here, we evaluate the ability of LLMs to compose their reasoning process over parts of the task (program statements in this case) in new ways, as opposed to identifying and ignoring irrelevant information.

\subsection{Robustness of Code LLMs}
Testing the robustness of code LLMs to semantics-preserving transformations is a widely studied problem in the literature~\citep{gao2023discrete,sarker2024syntactic,yang2024robustness,hooda2024large,wang2023recode}. These works utilize techniques such as renaming variables, permuting statements without dependencies, branch rewriting, and adding dead code to evaluate how an LLM's performance is affected by these changes, which are designed specifically to preserve the semantics of the code. Therefore, this line of research measures how LLMs generalize to new representations of the same functionality. In contrast, our mutation mechanism modifies programs with the explicit goal of changing semantics while preserving overall code complexity, enabling research that measures generalization to new functionality expressed with similar complexity. Most previous robustness works also focus on code generation, while we focus on the task of code reasoning.

\subsection{Evaluating Different Aspects of Code Reasoning}
Researchers have proposed multiple benchmarks to measure the ability of LLMs to reason about different aspects of program behavior. \citet{gu2024cruxeval} evaluate the code execution capabilities of LLMs on a dataset comprising simple Python functions generated by Code Llama~\citep{roziere2023code}. CodeMind~\citep{liu2024codemind} augments the previous work's framework to obtain a benchmark with additional programming languages and tasks such as dependent execution and specification reasoning. REval~\citep{chen2025reasoning} proposes to not only measure LLMs' abilities to predict program output, but also intermediate states of the program's execution. Our work is orthogonal and complementary to this line of research. We benchmark code execution as it is the most fundamental code reasoning task, but our approach of obtaining new programs (i.e., DSL sampling, mutation) and comparing performance across different distributions of code is directly applicable to other settings. \citet{li2024mutation} use mutation testing to explore whether LLMs can detect inconsistencies between a program's natural language description and mutated code. This approach tests if LLMs can match natural language with a program, but the goal is not to determine whether LLMs can actually reason about code behavior.

%% file: dsl_mutation.tex
\section{Obtaining New Programs}
\subsection{DSL Sampling} \label{sec:dsl_sampling}
Our first approach to acquiring programs for benchmarking LLM generalization in code reasoning is to (1) define a functional DSL and a set of syntactic constraints and (2) sample programs from the context-free grammar (CFG) compiled from the DSL and constraints. We describe this procedure in detail below. A DSL is defined by a set of primitives, together with the type and semantics of each primitive. Syntactic constraints include a diverse set of properties that one may specify to limit the space of possible programs, e.g., maximum depth (of the abstract syntax tree) and the type of the program. Together, the DSL and the set of syntactic constraints are compiled to a CFG. For details on this procedure, we refer the reader to \citet{fijalkow2022scaling}. To sample a program, we first assign a probability to each production in the CFG. Starting from the initial non-terminal node, we sample a production according to the probability distribution at that node; then, we recursively sample productions until reaching a terminal node. Finally, a sampled program can be converted into any common programming language with a transpiler.

We note two properties about this technique. First, since the resulting CFG defines a combinatorially large space of programs, it is unlikely for any sampled program to have been in the training data. Second, the difficulty of the sampled programs can be controlled by varying the syntactic constraints, e.g., program depth and number of function parameters.

Concretely, we construct a dataset by implementing a list-processing DSL containing common list transformations, conditionals, and maps, then translating sampled programs into imperative Python code. We focus on list-processing programs because they comprise an important class of widely used functions. We present details on this dataset in Section~\ref{sec:dsl_list}.

\subsection{Program Mutation} \label{sec:mutation}
\begin{algorithm}
\caption{Mutated Dataset Creation}
\label{alg:mutation}

\begin{algorithmic}[1]
\Statex \textbf{Input:} Original dataset $D$
\Statex \textbf{Output:} Mutated dataset $D'$
\Procedure{MutatePrograms}{$D$}
\State $D' \gets \varnothing$
\For{$(P, x) \in D$} \label{algline:loop_start}
    \State $S \gets \mathit{Mutate}(P)$ \label{algline:mutate} \Comment{Generate all mutants}
    \For{$P' \in S$} \label{algline:filter_start} \Comment{Remove invalid mutants}
        \If{\textbf{not} $\left( \mathit{IsExecutable}(P', x) \textbf{ and } \mathit{HasDifferentOutputs}(P, P', x) \right)$}
            \State $S \gets S \setminus \{P'\}$
        \EndIf
    \EndFor \label{algline:filter_end}
    \If{$S \neq \varnothing$} \label{algline:empty_mutation_set} \Comment{Select a mutant}
        \State $P' \gets \mathit{MostSimilarCoverage}(P, S, x)$ \label{algline:coverage}
        \State $D' \gets D' \cup \{(P', x)\}$ \label{algline:add_to_dataset}
    \EndIf
\EndFor \label{algline:loop_end}
\State \Return $D'$
\EndProcedure
\end{algorithmic}
\end{algorithm}

Mutation testing~\citep{papadakis2019mutation,jia2010analysis} applies local \emph{mutation operators} to programs. Originally developed to assess test suite quality, we use mutation to obtain new programs with different behaviors as a means to evaluate LLM generalization in code reasoning. At a high level, for each program-input pair in a given (original) dataset, we create a mutant that executes without error and produces a different output from the original program.

\begin{wraptable}{r}{0.45\textwidth}
    \centering
    \caption{Mutation operators and examples.}
    \label{tab:mutation_ops}

    \begin{tabular}{ll}
        \toprule
        Mutation Type & Example \\
        \midrule
        Arithmetic Operator & \texttt{a + b} $\to$ \texttt{a - b} \\
        Relational Operator & \texttt{a < b} $\to$ \texttt{a <= b} \\
        Logical Operator & \texttt{a and b} $\to$ \texttt{a or b} \\
        Keyword & \texttt{continue} $\to$ \texttt{break} \\
        Numerical Literal & \texttt{1} $\to$ \texttt{0} \\
        \bottomrule
    \end{tabular}
\end{wraptable}

Algorithm~\ref{alg:mutation} presents our mutation procedure. This procedure takes as input a dataset $D$ consisting of tuples $(P, x)$, where $P$ is a program and $x$ is an input to the program. The algorithm loops over each program-input pair $(P, x) \in D$ (Lines~\ref{algline:loop_start}--\ref{algline:loop_end}) and performs the following steps in each iteration. First, it modifies $P$ by applying the mutation operators shown in Table~\ref{tab:mutation_ops}. It tries all mutations separately, producing a set $S$ consisting of all programs $P$ with one mutation applied (Line~\ref{algline:mutate}). Then, it filters $S$ so that $S$ only consists of programs that (1) execute without error on the input $x$ and (2) produce a different output than the original program, i.e., $P(x) \neq P'(x)$ (Lines~\ref{algline:filter_start}--\ref{algline:filter_end}). If $S$ is now empty, it skips to the next iteration without adding to $D'$. Otherwise, it selects the mutated program $P'$ that has the most similar line coverage on $x$ compared to $P$; if there are multiple, it selects one uniformly at random (Line~\ref{algline:coverage}). This check ensures that the mutation approximately preserves the complexity of the problem and does not make the program trivial (e.g., returning on the second line after a condition on the first line becomes always true). This program $P'$ and the input $x$ are then added to $D'$ (Line~\ref{algline:add_to_dataset}). Some problems in $D$ may not have any mutant that results in valid code that runs without error and produces a different output (i.e., those with empty $S$ on Line~\ref{algline:empty_mutation_set}). In this case, we remove these problems from the original dataset $D$ to maintain a one-to-one correspondence between the problems in the original and mutated datasets. We apply the mutation procedure described above to each of the three datasets used in our evaluation (Section~\ref{sec:datasets}).

%% file: benchmarks.tex
\section{Code Reasoning Benchmark Sets} \label{sec:datasets}
\begin{figure}[t]
\centering

\begin{minipage}[t]{0.325\textwidth}
  \centering
  \vspace{0.5pt}
  \includegraphics[width=\linewidth]{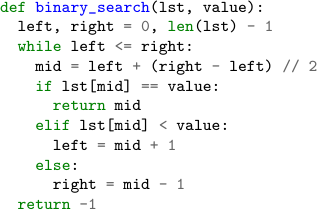}
\end{minipage}
\hfill
\tikz[baseline]{\draw[dotted, line width=0.5pt] (0,0) -- (0,-3.9);}
\hfill
\begin{minipage}[t]{0.23\textwidth}
  \centering
  \vspace{0pt}
  \includegraphics[width=\linewidth]{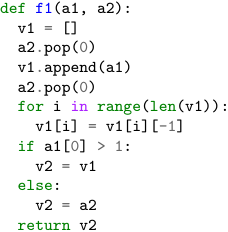}
\end{minipage}
\hfill
\tikz[baseline]{\draw[dotted, line width=0.5pt] (0,0) -- (0,-3.9);}
\hfill
\begin{minipage}[t]{0.41\textwidth}
  \centering
  \vspace{0pt}
  \includegraphics[width=\linewidth]{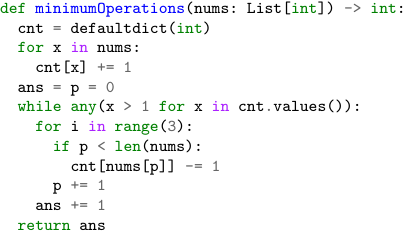}
\end{minipage}

\caption{Example programs from the LLM-List (left), DSL-List (center), and LeetCode (right) datasets.}
\label{fig:program_examples}
\end{figure}

We describe how we create the three benchmark sets used in our evaluation. Each dataset comprises a collection of program-input pairs, with the ground truth being the corresponding output. We ensure that each program is deterministic and thus has only one possible output. Fig.~\ref{fig:program_examples} shows an example program from each of our datasets. Fig.~\ref{fig:program_examples_appendix} in Appendix~\ref{sec:example_programs} contains extra examples of problems from each dataset (along with their mutated versions).

\subsection{DSL-List} \label{sec:dsl_list}
We define a list-processing DSL with the primitives and their associated types shown in Fig.~\ref{fig:list_dsl}. The semantics of each primitive is standard and corresponds to their naming. We choose these primitives because they are expressive enough to represent a rich class of list transformations while being sufficiently simple so that sampled programs are likely to be valid (i.e., do not produce runtime errors). We apply a variety of constraints designed to exclude trivial constructs from the sampled programs (e.g., expressions like \texttt{0 == 0}); these are detailed in Appendix~\ref{sec:dsl_constraints}. To include programs of different types and sizes in our dataset, we specify the programs to be of type \texttt{List(int)} $\rightarrow$ \texttt{List(int)} or \texttt{List(int)} $\rightarrow$ \texttt{List(int)} $\rightarrow$ \texttt{List(int)} (i.e., list transformations that take in one or two inputs) and to be of depth 4 or 5. We assign probabilities to the productions in the compiled CFG so that \texttt{if} and \texttt{map} have a weight of 5, \texttt{extend} has a weight of 0.05, and all other primitives have a weight of 1. We implement our language and sampler in Python by adapting the framework from \citet{fijalkow2022scaling}. We translate each sampled DSL program to imperative Python code using Algorithm~\ref{alg:translation} in Appendix~\ref{sec:dsl_translation}.

When sampling each program, we also sample three inputs according to the program's type signature. Lists in these inputs consist of three to five elements, with each element being an integer in $[0, 5]$; the list lengths and elements are uniformly sampled with replacement. We consider a program to be valid if it executes without error on all three inputs. 

To create our dataset, we sample 1000 valid programs for each combination of program type and program depth. For each type signature, we select a random subset of 50 programs such that there are 10 programs each when binning by lines of code from 4 to 24 (with a bin width of 4). In total, we obtain 100 programs with three inputs for each program.

\begin{figure}[t]
    \small
    \centering

    \hfill
    \begin{minipage}[t]{0.37\textwidth}
        \begin{enumerate}[leftmargin=0pt]
            \item[] \textbf{if:} \texttt{bool $\to$ t0 $\to$ t0 $\to$ t0}
            \item[] \textbf{map:} \texttt{(t0 $\to$ t1) $\to$ L(t0) $\to$ L(t1)}
            \item[] \textbf{empty:} \texttt{L(t0)}
            \item[] \textbf{append:} \texttt{t0 $\to$ L(t0) $\to$ L(t0)}
            \item[] \textbf{extend:} \texttt{L(t0) $\to$ L(t0) $\to$ L(t0)}
        \end{enumerate}
    \end{minipage}
    \begin{minipage}[t]{0.3\textwidth}
        \begin{enumerate}[leftmargin=0pt]
            \item[] \textbf{init:} \texttt{L(t0) $\to$ L(t0)}
            \item[] \textbf{tail:} \texttt{L(t0) $\to$ L(t0)}
            \item[] \textbf{length:} \texttt{L(t0) $\to$ int}
            \item[] \textbf{index:} \texttt{int $\to$ L(t0) $\to$ t0}
            \item[] \textbf{$\bm{==}$:} \texttt{int $\to$ int $\to$ bool}
        \end{enumerate}
    \end{minipage}
    \begin{minipage}[t]{0.3\textwidth}
        \begin{enumerate}[leftmargin=0pt]
            \item[] \textbf{$\bm{<}$:} \texttt{int $\to$ int $\to$ bool}
            \item[] \textbf{$\bm{>}$:} \texttt{int $\to$ int $\to$ bool}
            \item[] \textbf{\&\&:} \texttt{bool $\to$ bool $\to$ bool}
            \item[] \textbf{$\bm{\vert \vert}$:} \texttt{bool $\to$ bool $\to$ bool}
            \item[] \textbf{!:} \texttt{bool $\to$ bool}
        \end{enumerate}
    \end{minipage}
    \hfill

    \caption{Domain-specific language for the DSL-List dataset. \texttt{L} denotes \texttt{List}, and \texttt{t0} and \texttt{t1} are polymorphic types. We allow integers in the range $[-1, 5]$.}
    \label{fig:list_dsl}
\end{figure}

\subsection{LLM-List}
We obtain the LLM-List dataset in a three-step process: brainstorming, code generation, and input generation. All prompts described below are shown in Appendix~\ref{sec:llm_list_details}.

We first use GPT-4o to brainstorm 100 common list functions in Python. Specifically, we instruct the model to come up with functions that take in a list of integers as one of the parameters and do not contain random operations, substantial floating-point operations, or additional imports. The model produces a numbered list of function headers (e.g., \texttt{remove(lst, value)}) and their natural language descriptions. Examples of brainstormed operations in this phase include \texttt{merge}, \texttt{slice}, and \texttt{intersection}. Afterward, we add manually specified headers and descriptions for ten sorting algorithms and two search algorithms to the brainstormed list, obtaining a collection of function headers and descriptions for 112 common list operations.

Next, we instruct GPT-4o to generate the code for the brainstormed operations. We provide the model with each function's header and description, and instruct it to implement a deterministic function with limited usage of built-ins, while ensuring the logic in the function is as explicit as possible. The model generates well-structured, readable programs with meaningful variable names.

The generated list programs have a variety of type signatures. We use GPT-4o to generate three test inputs for each program and use these inputs in our dataset. The prompt specifies that each input should produce no errors when executed, should not include floating-point numbers, and that lists should only contain a few elements, but not be empty. We execute the generated inputs and regenerate them in case any of these constraints are violated. In total, we obtain 112 programs with three inputs for each program.

\subsection{LeetCode}
LeetCode is a dataset of human-submitted solutions to competitive programming problems that are collected from the LeetCode contest website~\citep{leetcode}. LiveCodeBench~\citep{jain2025livecodebench} also used LeetCode contest solutions to evaluate LLMs on code execution prediction. However, their dataset only covers the time period between May 2023 and November 2023 (inclusive), which is insufficient for evaluating newer models whose knowledge cutoffs are more recent. Our dataset consists of solutions submitted between May 2023 and January 2025 (inclusive). We divide this dataset into two halves: one containing programs from May 2023 to August 2023 (inclusive), before the knowledge cutoffs of all the LLMs we consider; and one containing programs during or after August 2024, after the cutoffs of all the LLMs we consider. 

LeetCode runs six contests a month (including weekly and biweekly contests), and each contest has four questions. Consistent with LiveCodeBench, we select the subset of these questions that overlap with their code generation dataset. For each question, we collect up to the top 20 ranked solutions written in Python3. We exclude solutions that do not pass all of the public test cases and deduplicate solutions with identical abstract syntax trees. We also apply several compile-time and runtime filters to ensure that problems in our dataset are reasonable. We exclude problems whose code is outside the range of 100 to 800 characters or have more than 1000 bytecode operations (excluding imports) and apply other filters identical to LiveCodeBench. We note that LiveCodeBench excluded problems whose code is outside the range of 100 to 500 characters or have more than approximately 500 bytecode operations (excluding imports). We increased the difficulty of our problems because of the increased code reasoning capabilities of more recent LLMs. To ensure that our dataset contains a diverse set of problems while remaining manageable in size, for each question, we include up to three solutions (each paired with an input from the public test cases) that are selected uniformly at random. In total, we assemble 184 and 190 problems before and after cutoff, respectively.

%% file: experiments.tex
\section{Experimental Methodology}
We detail the two sets of experiments that we run, followed by a discussion of our experimental setup.

\subsection{Execution Prediction} \label{sec:exec_pred}
\begin{figure}
    \centering
    \includegraphics[width=0.85\textwidth]{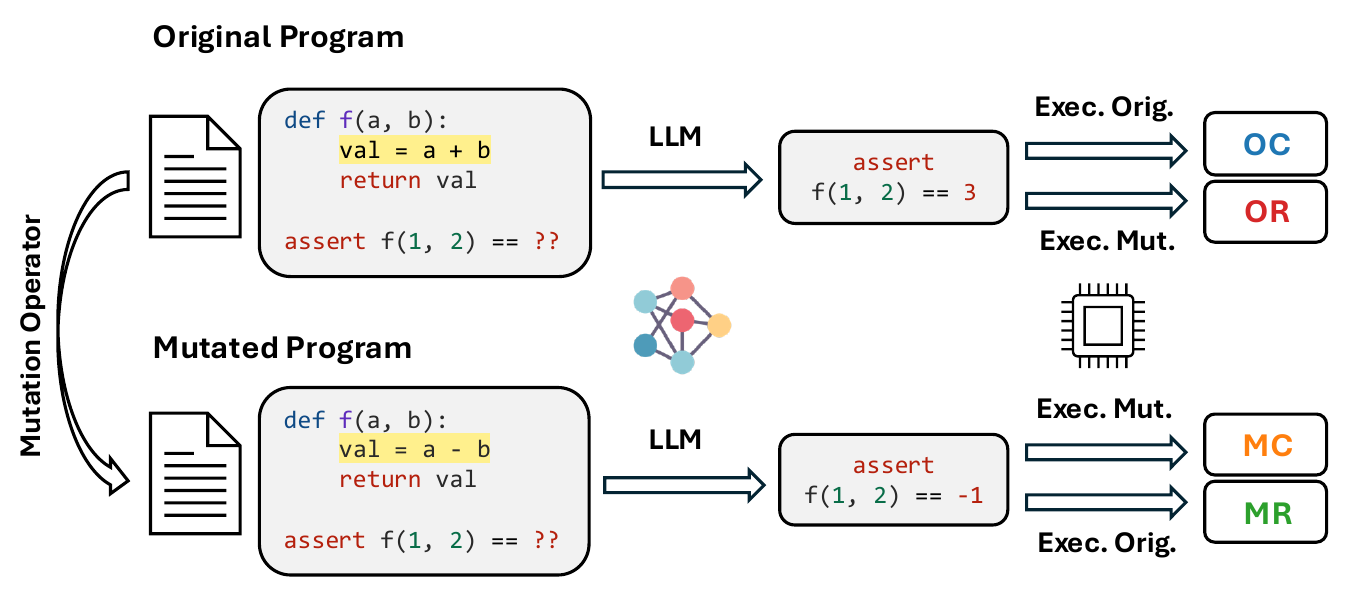}

    \caption{Overview of the execution prediction experiment. Given an original program $P$, its mutated version $P'$, and an input $x$, we instruct an LLM to predict the outputs of the original program $P(x)$ and the mutated program $P'(x)$. For the original program, we check if the model prediction correctly matches $P(x)$ (OC) or incorrectly matches $P'(x)$ (OR). For the mutated program, we check if the model prediction correctly matches $P'(x)$ (MC) or incorrectly matches $P(x)$ (MR).}
    \label{fig:exec_pred_overview}
\end{figure}

Our first set of experiments measures the ability of an LLM to, given a program and an input, correctly predict the output of the program when executed on the input. To instruct the models on this task, we adapt the prompt from \citet{jain2025livecodebench}. For models not specifically trained for reasoning (DeepSeek-Coder, Qwen2.5-Coder, and GPT models), we utilize chain-of-thought prompting~\citep{wei2022chain} along with few-shot prompting~\citep{brown2020language} to maximize the code reasoning capabilities of these models. In line with best practices, we do not provide any examples (i.e., use a zero-shot setting) for the reasoning models to allow them maximum flexibility in their reasoning chain. Our complete prompts are in Appendix~\ref{sec:exec_pred_prompts}.

Fig.~\ref{fig:exec_pred_overview} provides an overview of our experimental procedure. Each problem (i.e., program-input pair) in each dataset has both a version with the original program $P$ and another with the mutated program $P'$. Given an input $x$, we instruct the model to predict the outputs of the original and mutated programs in separate inference passes. Then, we measure:
\begin{itemize}
    \item \textbf{Original Correctness (OC):} Whether the LLM predicts $P(x)$ when given the original program.
    \item \textbf{Original Reversion (OR):} Whether the LLM reverts to (i.e., incorrectly predicts) $P'(x)$ when given the original program.
    \item \textbf{Mutation Correctness (MC):} Whether the LLM predicts $P'(x)$ when given the mutated program.
    \item \textbf{Mutation Reversion (MR):} Whether the LLM reverts to (i.e., incorrectly predicts) $P(x)$ when given the mutated program.
\end{itemize}
Recall in Section~\ref{sec:mutation} that we ensure $P(x) \neq P'(x)$ by construction; thus, reversion always indicates an incorrect prediction.

\subsection{Execution Choice} \label{sec:exec_choice}
We perform a second set of experiments, which measures the ability of an LLM to, given two programs and an input, first pick a program, then correctly predict its output when executed on the input. For each problem, we provide the original program and its corresponding mutated program as the two programs. We specifically instruct the model to pick whichever program it is more confident in reasoning about, without providing any information on the existence of mutations. The prompts we use are shown in Appendix~\ref{sec:exec_choice_prompts}. Then, we measure:
\begin{itemize}
    \item \textbf{Preference:} Whether the LLM chooses to reason about the original or the mutated program.
    \item \textbf{Correctness:} Whether the LLM correctly predicts the output for the program it chooses.
    \item \textbf{Reversion:} Whether the LLM predicts the output for the other program (that it does not choose).
\end{itemize}

\subsection{Experimental Setup}
\subsubsection{Benchmarked Large Language Models}
\begin{table}
    \centering
    \caption{Overview of the LLMs used in our evaluation.}
    \label{tab:model_overview}

    \begin{tabular}{llrrr}
        \toprule
        Model Family & Model Name & Size & \makecell[r]{Approximate\\Cutoff Date} & \makecell[r]{Release\\Date} \\
        \midrule
        \multirow{2}{*}{\makecell[l]{DeepSeek-\\Coder (DC)}} 
        & DC-7B & 7B & 2023-09 & 2023-11 \\
        & DC-33B & 33B & 2023-09 & 2023-11 \\
        \midrule
        \multirow{3}{*}{\makecell[l]{Qwen2.5-\\Coder (QC)}} 
        & QC-7B & 7B & 2024-07 & 2024-09 \\
        & QC-14B & 14B & 2024-07 & 2024-11 \\
        & QC-32B & 32B & 2024-07 & 2024-11 \\
        \midrule
        QwQ & QwQ-32B & 32B & 2024-08 & 2025-03 \\
        \midrule
        DeepSeek-R1 & DeepSeek-R1 & 671B & 2024-07 & 2025-01 \\
        \midrule
        \multirow{3}{*}{OpenAI}
        & GPT-4o-mini & - & 2023-10 & 2024-07 \\
        & GPT-4o & - & 2023-10 & 2024-05 \\
        & o3-mini & - & 2023-10 & 2025-01 \\
        \bottomrule
    \end{tabular}
\end{table}

We benchmark a representative set of 10 large language models that includes state-of-the-art open-access and closed-access models over the past year. These models cover five model families, different model sizes within the same family, and both traditional and recent reasoning LLMs. Our open-access models include DeepSeek-Coder-$\{$33, 7$\}$B (state-of-the-art at the beginning of 2024) and Qwen2.5-Coder-$\{$32, 14, 7$\}$B (state-of-the-art toward the end of 2024), as well as QwQ-32B and DeepSeek-R1, the newest class of reasoning models trained with reinforcement learning to perform extensive chain-of-thought reasoning. We use the instruction-tuned versions of DeepSeek-Coder and Qwen2.5-Coder. Our closed-access LLMs include OpenAI models that span traditional and new reasoning models: GPT-4o-mini, GPT-4o, and o3-mini. Table~\ref{tab:model_overview} presents an overview of these models, including their approximate knowledge cutoff and release dates. The cutoff date means that the model was only trained on data that was available prior to the displayed month. Table~\ref{tab:model_checkpoints} in Appendix~\ref{sec:model_identifiers} lists the specific identifiers for each model.

\subsubsection{Generation Parameters}
Following \citet{jain2025livecodebench}, we use a temperature of 0.2 and top-$p$ value of 0.95 to sample responses from traditional models (DeepSeek-Coder, Qwen2.5-Coder, and GPT models). For QwQ and DeepSeek-R1, we use the recommended temperature setting of 0.6 to prevent endless repetitions or incoherent outputs, along with a top-$p$ of 0.95. For o3-mini, parameters like temperature and top-$p$ are not supported. Instead, a parameter called reasoning effort controls how much thinking the model conducts; we set this parameter to high. We constrain the maximum generation length to 4096 tokens for DeepSeek-Coder and Qwen2.5-Coder, as we observe that outputs above this length contain endless looping; we do not constrain the generation length for other models and allow them to perform maximal chain-of-thought thinking. All other parameters are set at their default values.

\subsubsection{Inference Setup}
To obtain our results on DeepSeek-Coder, Qwen2.5-Coder, and QwQ, we download the models from \citet{huggingface} and use vLLM~\citep{kwon2023efficient} to perform efficient inference on a cluster of A100-80GB GPUs. For DeepSeek-R1 and the OpenAI models, we use the respective APIs of the two companies.

\subsubsection{Evaluation Metrics} \label{sec:metrics}
For the execution prediction experiments, we generate five responses for each problem and compute the pass@1 metric~\citep{kulal2019spoc,chen2021evaluating} for OC, OR, MC, and MR (explained in Section~\ref{sec:exec_pred}). This metric measures the probability that a model produces the answer that fits the corresponding criterion in one generation, which is estimated by dividing the number of fitting answers by five. When reporting reversion rates, we exclude problems whose output is a Boolean variable; this choice makes the reversion metric more meaningful since it does not include cases where a model simply produces an incorrect result.

For the execution choice experiments, we conduct two runs for each problem and swap the order of the original and mutated programs to mitigate any bias a model may have for the ordering of the programs. To quantify preference (see Section~\ref{sec:exec_choice}), we assign a value of $0$ if a model chooses the mutated program and a value of $1$ if it chooses the original program. In this setting, we use OC to denote the percentage of correct responses when a model chooses to reason about the original program; OR denotes the percentage of responses that predict the output of the mutated program when a model chooses to reason about the original program. MC and MR are defined analogously. Similar to the execution prediction setting, we exclude problems with Boolean outputs when calculating reversion.

%% file: results.tex
\section{Results}
We present the results from our experiments. In each section, we first provide an overview of the tables and figures that contain the results. Then, we discuss the execution prediction results, execution prediction results as a function of program complexity, and execution choice results in separate subsections. In each subsection, we highlight important trends in the results and discuss their implications.

\subsection{List Datasets}
\begin{table}
\centering
\caption{Execution prediction and choice results on list datasets.}
\label{tab:list_res}
{
\setlength{\tabcolsep}{3pt}
\begin{tabular}{llrrrrp{0.6cm}rrrrr}
\toprule
\multirow[c]{2}{*}{Dataset} & \multirow[c]{2}{*}{Model} & \multicolumn{4}{c}{Execution Prediction} & & \multicolumn{5}{c}{Execution Choice} \\
\cmidrule(lr){3-6} \cmidrule(lr){8-12}
& & OC ($\uparrow$) & MC ($\uparrow$) & OR ($\downarrow$) & MR ($\downarrow$) & & Pref & OC ($\uparrow$) & MC ($\uparrow$) & OR ($\downarrow$) & MR ($\downarrow$) \\
\midrule
\multirow[c]{10}{*}{\makecell[l]{LLM-\\List}} & DC-7B & 85.4 & 31.6 & 3.1 & 55.5 &  & 79.0 & 81.7 & 18.6 & 5.0 & 64.1 \\
 & DC-33B & 91.7 & 37.3 & 2.3 & 54.0 &  & 83.0 & 85.3 & 33.3 & 3.8 & 44.6 \\
 & QC-7B & 94.1 & 58.7 & 1.7 & 29.7 &  & 90.9 & 90.3 & 35.7 & 3.4 & 56.2 \\
 & QC-14B & 97.1 & 70.3 & 0.8 & 22.4 &  & 94.4 & 98.2 & 42.3 & 0.3 & 52.2 \\
 & QC-32B & 99.6 & 73.4 & 0.0 & 26.1 &  & 96.8 & 99.3 & 40.0 & 0.0 & 69.2 \\
 & QwQ-32B & 100.0 & 95.7 & 0.0 & 3.0 &  & 98.9 & 100.0 & 100.0 & 0.0 & 0.0 \\
 & DeepSeek-R1 & 100.0 & 98.2 & 0.0 & 0.6 &  & 95.0 & 100.0 & 95.7 & 0.0 & 0.0 \\
 & GPT-4o-mini & 99.3 & 68.1 & 0.3 & 28.5 &  & 92.7 & 98.6 & 41.2 & 0.6 & 59.3 \\
 & GPT-4o & 100.0 & 80.1 & 0.0 & 16.0 &  & 98.1 & 99.8 & 100.0 & 0.3 & 0.0 \\
 & o3-mini & 100.0 & 99.9 & 0.0 & 0.0 &  & 97.4 & 100.0 & 100.0 & 0.0 & 0.0 \\
\midrule
\multirow[c]{10}{*}{\makecell[l]{DSL-\\List}} & DC-7B & 24.6 & 24.2 & 4.9 & 4.8 &  & 46.6 & 12.8 & 16.5 & 9.1 & 5.4 \\
 & DC-33B & 45.8 & 45.5 & 7.0 & 7.6 &  & 55.4 & 26.8 & 30.8 & 9.3 & 9.4 \\
 & QC-7B & 61.6 & 58.1 & 2.5 & 4.7 &  & 55.8 & 52.4 & 58.1 & 8.4 & 6.0 \\
 & QC-14B & 77.3 & 76.7 & 1.3 & 0.8 &  & 53.8 & 77.8 & 76.6 & 1.4 & 1.2 \\
 & QC-32B & 85.7 & 87.2 & 1.7 & 0.8 &  & 49.2 & 81.2 & 85.5 & 4.6 & 3.0 \\
 & QwQ-32B & 98.1 & 97.9 & 0.2 & 0.1 &  & 52.5 & 99.3 & 98.8 & 0.0 & 0.0 \\
 & DeepSeek-R1 & 97.7 & 98.9 & 0.5 & 0.0 &  & 48.4 & 99.6 & 99.3 & 0.0 & 0.0 \\
 & GPT-4o-mini & 76.4 & 77.4 & 1.4 & 1.4 &  & 50.4 & 70.0 & 74.9 & 4.5 & 3.0 \\
 & GPT-4o & 91.3 & 93.5 & 1.4 & 0.2 &  & 54.9 & 88.0 & 90.4 & 2.4 & 2.5 \\
 & o3-mini & 98.7 & 98.5 & 0.0 & 0.0 &  & 50.8 & 100.0 & 100.0 & 0.0 & 0.0 \\
\bottomrule
\end{tabular}
}
\end{table}

\begin{figure}
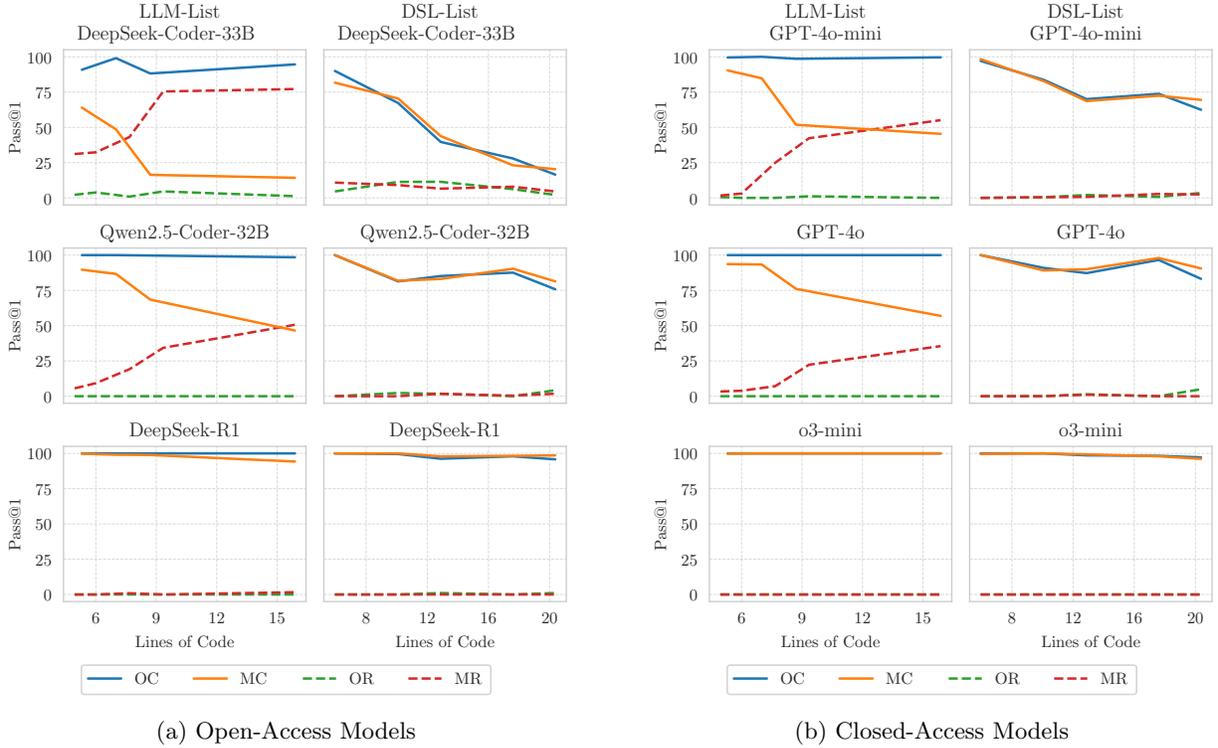

    \centering
    \begin{subfigure}[b]{0.48\textwidth}
        \centering
        \resizebox{\linewidth}{!}{\input{figures/list_open.pgf}}
        \caption{Open-Access Models}
    \end{subfigure}
    \hfill
    \begin{subfigure}[b]{0.48\textwidth}
        \centering
        \resizebox{\linewidth}{!}{\input{figures/list_closed.pgf}}
        \caption{Closed-Access Models}
    \end{subfigure}

    \caption{Execution prediction results on list datasets as a function of lines of code.}
    \label{fig:list_loc}
\end{figure}

Table~\ref{tab:list_res} presents the results for the execution prediction and choice experiments on the LLM-List and DSL-List datasets. The table contains a row of results for each dataset-model combination; the first and second columns identify the dataset and the model, respectively. The following columns contain results on the two sets of experiments, with each entry consisting of the corresponding metric described in Section~\ref{sec:metrics}, averaged over all problems in the dataset. Fig.~\ref{fig:list_loc} presents the execution prediction results on the list datasets as a function of program complexity; we use lines of code as a simple proxy for program complexity. In each subfigure, the left and right columns show results on the LLM-List and DSL-List datasets, respectively, while each row shows results for a different model. Each graph contains four lines: the blue, orange, green, and red lines correspond to OC, MC (correctness metrics) and OR, MR (reversion metrics).

\subsubsection{Execution Prediction Results} 
In Table~\ref{tab:list_res}, the OC and MC columns under Execution Prediction present how often a model predicts the correct output when given the original and mutated programs, respectively (higher is better). The OR and MR columns present how often a model, when given the original and mutated programs, respectively, predicts the output of the other program (lower is better).

On LLM-List, correctness is generally high for the original problems. Earlier open-access models (DC) exhibit a sharp drop in correctness (around 54 percentage points) from the original to mutated problems. Later traditional open-access models (QC) still exhibit significant drops (between 26--35 points), but to a much lesser degree than DC. There is also a gap between the original and mutated scores for GPT models, with the performance of GPT-4o-mini being comparable to QC-14B and GPT-4o being better than QC-32B. In the same family, correctness increases with model size (especially on mutated problems). The latest reasoning models (QwQ, DeepSeek-R1, o3-mini) show very little degradation in correctness from original to mutated problems (at most a 4-point drop); notably, o3-mini achieves essentially perfect performance. OR is close to zero for all models. DC models exhibit substantial reversion on mutated programs (above 50\%); QC models still exhibit high reversion but to a much lesser degree than DC (from 22--30\%). GPT-4o-mini has a reversion rate close to that of QC-7B, while GPT-4o has the lowest reversion compared to the other traditional LLMs, at around 16\%. In contrast, the latest reasoning models exhibit very low reversion (at most 3\%), with o3-mini showing no reversion. 

On DSL-List, the correctness rates for the original and mutated problems are comparable for each model. DC models achieve very low scores (24--46\%), while QC models are notably better in comparison (61--87\%). Again, GPT-4o-mini's performance is comparable to that of QC-14B, and GPT-4o is slightly better than QC-32B. In the same family, correctness increases significantly with model size. The latest reasoning models exhibit near-perfect performance. OR is comparable to MR for each model, and the reversion rate is in the single digits for all models.

These results highlight several important phenomena. They are consistent with the propositions that LLM-List programs are in-distribution and DSL-List programs are relatively out-of-distribution; also, mutation moves LLM-List programs out-of-distribution, while not significantly affecting the distribution of DSL-List programs. Furthermore, earlier models exhibit behavior consistent with significant pattern matching on in-distribution LLM-List problems, while the latest models exhibit behavior consistent with generalization. Earlier models have difficulty reasoning about relatively simple out-of-distribution DSL-List problems, while the latest models succeed at a near-perfect rate.

\subsubsection{Execution Prediction Results vs. Program Complexity}
To investigate how execution prediction ability varies with program complexity, we plot the results discussed above against lines of code for a subset of open- and closed-access models in Fig.~\ref{fig:list_loc}. The results for the other models are shown in Fig.~\ref{fig:list_loc_open_appendix} in Appendix~\ref{sec:results_appendix}. On LLM-List, correctness for original problems remains high regardless of program size; correctness for mutated problems drops and reversion increases as program size increases. This effect is extremely pronounced for earlier open-access models. For instance, on DC-33B, the gap between original and mutated problems is 80 points and reversion on mutated problems is 77\% for programs around 15 lines. Even state-of-the-art traditional LLMs like QC-32B and GPT-4o still exhibit a substantial original-mutation drop and high reversion at higher program complexity. On the other hand, reasoning models like DeepSeek-R1 and o3-mini demonstrate near-perfect correctness and low reversion across all complexities on both the original and mutated problems. On DSL-List, original and mutated scores are close to identical and reversion is negligible for each model. The correctness of traditional LLMs decreases with program size; DC scores rapidly decline, while QC and GPT scores decline at a slower rate. The newest reasoning models are highly accurate at all program sizes.

These results are consistent with traditional LLMs somewhat generalizing on smaller programs, but their reasoning capabilities quickly falling with increasing problem complexity. In contrast, the latest reasoning LLMs are competent at reasoning across all program sizes, showing only a very slight drop in performance as problem complexity increases.

\subsubsection{Execution Choice Results}
In Table~\ref{tab:list_res}, the Pref column under Execution Choice presents the percentage of problems for which the LLM chooses to predict results for the original instead of the mutated code. We emphasize that (as described in Section~\ref{sec:exec_choice}), there is no mention of original vs. mutated code in our prompt, and a model is only instructed to choose whichever program it is more confident in reasoning about. The OC and MC columns present how often a model predicts the correct output when choosing to reason about the original and mutated programs, respectively. The OR and MR columns present how often a model predicts the output of the other program when choosing to reason about the original and mutated programs, respectively.

On LLM-List, all models consistently choose the original program. The DC models choose the original code at around 80\% frequency, while all other models select the original code at over 90\% frequency. Original correctness is generally high and comparable to that of execution prediction. All traditional models with the exception of GPT-4o experience a large drop in correctness when choosing to reason about the mutated programs. Notably, for these models, MC in execution choice is much lower than MC in execution prediction. On the other hand, GPT-4o and the reasoning models achieve very high MC (96\% for DeepSeek-R1 and 100\% for the others). OR is negligible for all models. All traditional models except GPT-4o exhibit substantial reversion when choosing to reason about the mutated programs. This phenomenon is especially pronounced on the QC models and GPT-4o-mini, which revert at about twice the rate compared to execution prediction. In stark contrast, GPT-4o and the reasoning models do not revert at all.

On DSL-List, all models choose the original program roughly half of the time. The correctness scores when choosing to reason about the original and mutated programs are comparable for each model. The correctness scores of the DC models are noticeably lower compared to the execution prediction experiments; we attribute this drop to their difficulty in reasoning correctly about the second program in the prompt in this scenario. The performances of other models are comparable across the prediction and choice experiments. Reversion is low both when choosing to reason about the original and mutated programs for all models.

These results highlight the introspection capabilities of LLMs, since they can recognize in-distribution programs as those they are more confident in reasoning about --- with the models' preferences for original LLM-List programs generally increasing with their reasoning capabilities. On LLM-List, earlier and less capable models generally exhibit much lower correctness and higher reversion when choosing to reason about the mutated programs, further accentuating their reliance on pattern matching. In contrast, the latest reasoning models achieve high correctness for all problems despite overwhelmingly selecting the original programs when given the choice. This trend is consistent with mutation indeed moving in-distribution programs out-of-distribution, and also with these models exhibiting strong reasoning abilities that generalize to new problems.

\subsection{LeetCode Dataset}
\begin{table}
\centering
\caption{Execution prediction and choice results on LeetCode.}
\label{tab:leetcode_res}
{
\setlength{\tabcolsep}{3pt}
\begin{tabular}{llrrrrp{0.6cm}rrrrr}
\toprule
\multirow[c]{2}{*}{\makecell[l]{Time\\Period}} & \multirow[c]{2}{*}{Model} & \multicolumn{4}{c}{Execution Prediction} & & \multicolumn{5}{c}{Execution Choice} \\
\cmidrule(lr){3-6} \cmidrule(lr){8-12}
& & OC ($\uparrow$) & MC ($\uparrow$) & OR ($\downarrow$) & MR ($\downarrow$) & & Pref & OC ($\uparrow$) & MC ($\uparrow$) & OR ($\downarrow$) & MR ($\downarrow$) \\
\midrule
\multirow[c]{10}{*}{\makecell[l]{Before\\Cutoff}} & DC-7B & 29.8 & 16.1 & 6.8 & 18.5 &  & 60.7 & 36.7 & 20.1 & 10.5 & 19.0 \\
 & DC-33B & 46.8 & 24.0 & 10.6 & 29.8 &  & 63.6 & 39.1 & 24.0 & 7.9 & 17.6 \\
 & QC-7B & 59.7 & 48.9 & 6.0 & 14.0 &  & 73.6 & 55.7 & 38.3 & 7.8 & 33.3 \\
 & QC-14B & 74.9 & 68.7 & 2.8 & 12.1 &  & 86.3 & 72.2 & 59.2 & 5.1 & 19.0 \\
 & QC-32B & 80.4 & 68.0 & 4.1 & 18.3 &  & 93.3 & 78.4 & 50.0 & 3.1 & 31.8 \\
 & QwQ-32B & 97.9 & 95.8 & 0.1 & 2.0 &  & 95.3 & 97.9 & 100.0 & 0.0 & 0.0 \\
 & DeepSeek-R1 & 99.3 & 95.8 & 0.0 & 3.6 &  & 96.4 & 100.0 & 100.0 & 0.0 & 0.0 \\
 & GPT-4o-mini & 83.2 & 69.3 & 2.5 & 13.9 &  & 78.2 & 77.1 & 59.0 & 3.7 & 23.6 \\
 & GPT-4o & 91.2 & 77.9 & 1.4 & 12.2 &  & 93.6 & 85.1 & 82.6 & 1.4 & 4.5 \\
 & o3-mini & 99.9 & 99.9 & 0.0 & 0.0 &  & 89.4 & 99.7 & 100.0 & 0.0 & 0.0 \\
\midrule
\multirow[c]{10}{*}{\makecell[l]{After\\Cutoff}} & DC-7B & 21.6 & 18.4 & 9.3 & 10.2 &  & 59.9 & 29.0 & 23.1 & 9.6 & 15.7 \\
 & DC-33B & 34.5 & 25.5 & 12.8 & 16.0 &  & 66.8 & 34.8 & 33.3 & 10.2 & 15.0 \\
 & QC-7B & 51.4 & 50.8 & 10.2 & 12.4 &  & 71.0 & 50.8 & 45.4 & 10.5 & 13.5 \\
 & QC-14B & 67.9 & 64.1 & 6.9 & 11.5 &  & 82.4 & 64.9 & 60.6 & 5.5 & 18.9 \\
 & QC-32B & 75.8 & 69.1 & 8.3 & 13.3 &  & 88.0 & 77.8 & 66.7 & 6.6 & 26.3 \\
 & QwQ-32B & 98.6 & 95.5 & 0.0 & 3.3 &  & 89.3 & 98.5 & 100.0 & 0.0 & 0.0 \\
 & DeepSeek-R1 & 99.5 & 98.5 & 0.4 & 1.5 &  & 87.7 & 99.4 & 100.0 & 0.0 & 0.0 \\
 & GPT-4o-mini & 75.4 & 67.9 & 4.9 & 9.2 &  & 73.3 & 65.3 & 57.0 & 10.7 & 14.1 \\
 & GPT-4o & 89.4 & 79.6 & 3.7 & 8.6 &  & 85.6 & 79.1 & 81.5 & 3.6 & 6.5 \\
 & o3-mini & 99.8 & 99.9 & 0.1 & 0.0 &  & 85.0 & 100.0 & 100.0 & 0.0 & 0.0 \\
\bottomrule
\end{tabular}
}
\end{table}

\begin{figure}
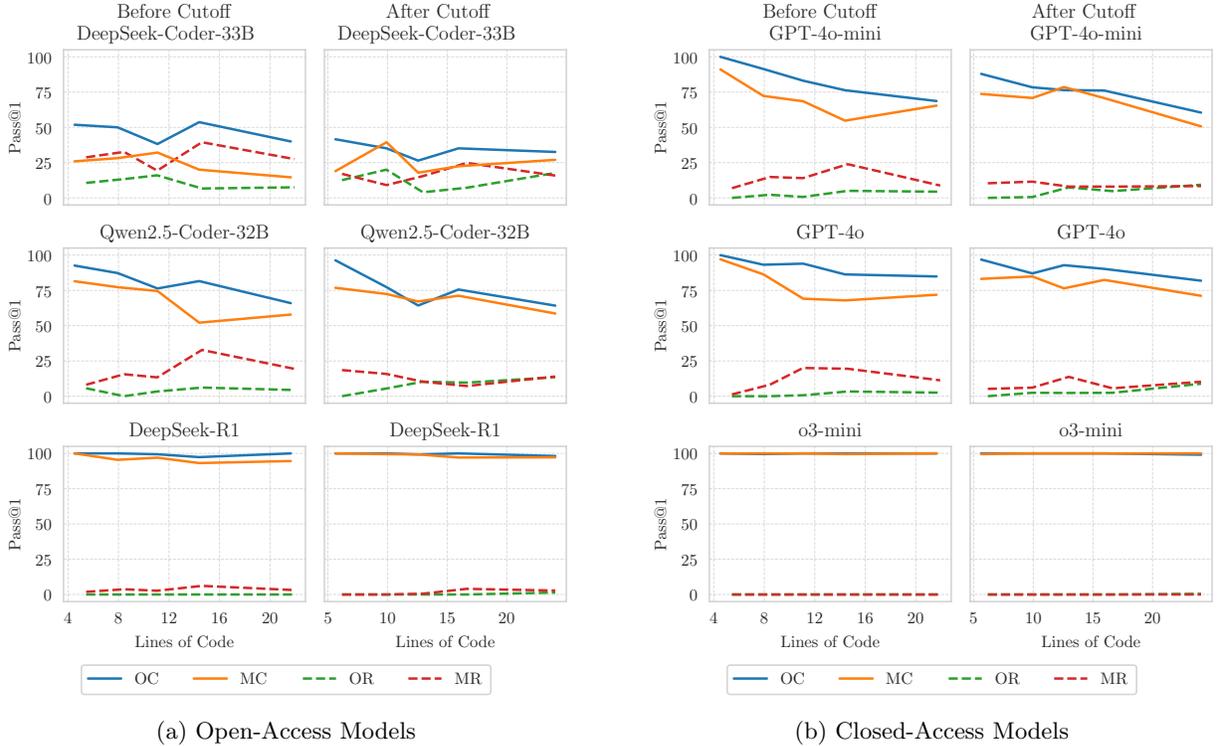

    \centering
    \begin{subfigure}[b]{0.48\textwidth}
        \centering
        \resizebox{\linewidth}{!}{\input{figures/leetcode_open.pgf}}
        \caption{Open-Access Models}
    \end{subfigure}
    \hfill
    \begin{subfigure}[b]{0.48\textwidth}
        \centering
        \resizebox{\linewidth}{!}{\input{figures/leetcode_closed.pgf}}
        \caption{Closed-Access Models}
    \end{subfigure}

    \caption{Execution prediction results on LeetCode as a function of lines of code.}
    \label{fig:leetcode_loc}
\end{figure}

Table~\ref{tab:leetcode_res} presents the results for the execution prediction and choice experiments on the LeetCode dataset, separated into results before and after cutoff. Fig.~\ref{fig:leetcode_loc} presents the execution prediction results on the LeetCode dataset as a function of lines of code. In each subfigure, the left and right columns, respectively, show results before and after cutoff. The layout of these results is otherwise identical to those of the list datasets.

\subsubsection{Execution Prediction Results}
Observing the results before cutoff, earlier open-access models (DC) struggle to correctly reason about code execution even on original problems (30--47\%); later traditional open-access models (QC) improve significantly but only up to around 80\%. The performance of GPT-4o-mini is comparable with that of QC-32B, while GPT-4o is noticeably better at around 91\%. In the same family, OC significantly increases with model size. The latest reasoning models are extremely capable, with OC in the range of 98--100\%. DC models exhibit a sharp drop in correctness (between 14--23 points) from the original to the mutated problems. QC and GPT models still exhibit significant drops (between 6--12 points and around 13 points, respectively), but to a lesser degree. In the same family, MC generally increases with model size. However, QC-14B and QC-32B attain similar MC; thus, their code reasoning capabilities on these problems may be more similar than their difference in OC suggests. QwQ and DeepSeek-R1 show slight degradation in correctness (at most a 4-point drop from original to mutated), while o3-mini remains essentially perfect. For all traditional LLMs, MR is significantly higher than OR by 10--20 points. All of these LLMs exhibit substantial reversion on mutated programs from around 12\% for GPT-4o and QC-14B to almost 30\% for DC-33B. The latest reasoning models exhibit little reversion, with o3-mini showing no reversion at all.

For the results after cutoff, traditional LLMs exhibit a noticeable drop in OC (compared to problems before cutoff); this drop is larger for DC models (8--12 points), but still present for QC models (5--8 points) and GPT models (2--8 points). This consistent drop indicates that some of the performance of these models on problems before cutoff is likely attributable to memorization. Therefore, benchmarking these models on problems after cutoff reduces the influence of data contamination on evaluation results. However, for traditional LLMs, there still exists a sizable gap between correctness on original and mutated problems after cutoff. For example, GPT-4o has a 10-point gap between OC and MC on problems after cutoff. This result (especially given the similar OC values before and after cutoff for GPT-4o) highlights that the strategy of evaluating models on problems after cutoff may be insufficient for truly measuring their out-of-distribution code reasoning capabilities. We posit that this is because certain patterns in the code remain similar for problems before and after cutoff. Conversely, our mutation strategy is effective at creating programs with out-of-distribution coding patterns. The correctness scores of reasoning models on problems before and after cutoff are comparable, with o3-mini still being near-perfect on both the original and mutated problems. Traditional LLMs exhibit lower reversion on mutated problems, which is consistent with reduced memorization on problems after cutoff. However, MR is still typically higher than OR (by 1--5 points). This bias toward original programs reveals that these models still exhibit some degree of pattern matching (toward common programming patterns), even on problems they have likely not seen before. The latest reasoning models exhibit reversion close to zero, similar to problems before cutoff.

These results highlight the importance of using techniques to create out-of-distribution problems (e.g., mutation) beyond using data cutoffs. They also continue to support the hypothesis that while traditional models (especially earlier ones) rely on memorization and pattern matching, the latest reasoning models exhibit strong generalization abilities on code execution prediction in comparison.

\subsubsection{Execution Prediction Results vs. Program Complexity}
We plot the results discussed above against lines of code for a subset of open- and closed-access models in Fig.~\ref{fig:leetcode_loc}. The results for other models are shown in Fig.~\ref{fig:leetcode_loc_open_appendix} in Appendix~\ref{sec:results_appendix}. We first discuss trends on the traditional LLMs. Both before and after cutoff, correctness on original and mutated problems generally decreases as program size increases. For a given complexity, correctness on problems before cutoff is higher for less capable models (e.g., DC-33B, GPT-4o-mini) while correctness before and after cutoff are comparable for more capable models (e.g., QC-32B, GPT-4o). The gap between original and mutated correctness is generally smaller for problems after cutoff. Before cutoff, reversion on mutated problems is generally lower for smaller programs and higher for larger programs; however, after cutoff, MR does not meaningfully change with program size. For a given complexity, reversion on mutated problems is higher before cutoff (especially for less capable models). Reasoning models like DeepSeek-R1 and o3-mini demonstrate near-perfect performance on both original and mutated problems and low reversion across all programs.

These results are consistent with the code reasoning abilities of traditional LLMs decreasing as program size increases on LeetCode problems. Also, for more capable models, their performance on problems before and after cutoff are highly similar, while the mutated problems challenge their generalization capabilities to a greater degree. The latest reasoning models can reason highly accurately for all program sizes that we consider, on both original and mutated problems.

\subsubsection{Execution Choice Results}
Both before and after cutoff, all models consistently prefer to reason about the original program. Before cutoff, earlier models like DC choose the original code at around 60\% frequency, while later models select the original code at around 90\% frequency. After cutoff, the preferences toward original programs for each model are largely similar but slightly lower than before cutoff. 

Before cutoff, original correctness is generally comparable to and follows the same trend as in the execution prediction experiments. All traditional models except GPT-4o experience a large drop in correctness when choosing to reason about the mutated programs. In contrast, GPT-4o attains MC comparable with OC, and the reasoning models attain 100\% MC. OR scores are low and comparable to those in the execution prediction experiments. All traditional models except GPT-4o exhibit substantial reversion (up to 33\%) when choosing to reason about the mutated programs. This phenomenon is especially striking for the QC models and GPT-4o-mini, which revert at up to twice the rate compared to execution prediction. GPT-4o only reverts 5\% of the time, while the reasoning models do not revert at all. The qualitative differences comparing performance results before and after cutoff are similar as in the execution prediction experiments.

These results once again indicate that the models (especially more capable ones) can clearly distinguish between original and mutated code, and prefer to reason about the original code. Low correctness and high reversion when choosing to reason about mutated programs are consistent with traditional LLMs' reliance on pattern matching. The ability of new reasoning models (especially o3-mini) to achieve near-perfect correctness for all problems is consistent with strong generalization capabilities on code reasoning.

%% file: conclusion.tex
\section{Conclusion}
Evaluating the generalization capabilities of LLMs on code reasoning is crucial toward understanding their ultimate impact on AI-assisted software engineering. We presented a range of techniques to obtain different classes of programs and an experimental methodology to evaluate LLM generalization by comparing their performance on these programs. Our results provided insights into the code reasoning capabilities of various LLMs on different classes of code and were consistent with the hypothesis that while prior models rely on pattern matching, the latest class of reasoning models exhibit strong generalization capabilities.

\section*{Acknowledgments}
This research was supported in part by the MIT–HPI Designing for Sustainability research program and by DARPA Grants HR001120C0191 and HR001124C0486. Any opinions, findings, and conclusions or recommendations expressed in this material are those of the authors and do not necessarily reflect the views of DARPA.

%% file: appendix.tex
\section{Dataset Details}
\subsection{DSL-List}
\subsubsection{Additional Constraints} \label{sec:dsl_constraints}
We apply the following syntactic constraints to all programs: (1) the first argument to a comparison operator must not be an integer literal (which avoids expressions like \texttt{0 == 0}), (2) the last argument to \texttt{extend}, \texttt{length}, and \texttt{map} must not be \texttt{empty}, (3) the integer literal \texttt{-1} is only used as an argument to \texttt{index}. We also optimize the compilation so that calling \texttt{index}, \texttt{init}, or \texttt{tail} on an empty list is disallowed (since these computations are guaranteed to lead to a runtime error), which makes program sampling much more efficient (as it rules out these programs by construction). The aforementioned rules are enforced when compiling the DSL to the CFG. Additional syntactic constraints are enforced during program sampling: (1) the same expression cannot appear on both sides of comparison or logical operators (which avoids expressions like \texttt{v1 == v1}), (2) a list cannot extend itself, (3) the same expression cannot be on both sides of a branch in a conditional, (4) a program must contain all function parameters. We also enforce the runtime constraint that a program must not produce the same output for all of its inputs. Programs that do not satisfy these constraints are discarded during sampling.

\subsubsection{Translation Algorithm} \label{sec:dsl_translation}
\begin{algorithm}
\caption{DSL to Python Program Translation}
\label{alg:translation}

\begin{algorithmic}[1]
\Statex \textbf{Input:} DSL program $D$
\Statex \textbf{Output:} Python program $P$
\Procedure{TranslateProgram}{$D$}
\State $T \gets \mathit{MakeAST}(D)$ \label{algline:make_ast}
\State $V \gets \mathit{SetVariables}(T)$ \label{algline:assign_vars}
\State $E \gets \mathit{SetExpressions}(T, V)$ \label{algline:assign_exprs}
\State $P \gets Init(T, E)$ \label{algline:init_vars}
\For{$T' \in \mathit{ReverseTopologicalOrder}(T)$} \label{algline:write_loop_start}
    \If{$\mathit{IsPartialNode}(T')$} \label{algline:partial_skip_start}
        \State \textbf{continue}
    \EndIf \label{algline:partial_skip_end}
    \If{$\mathit{IsMapNode}(T')$}
        \State $P \gets P \cdot^* ``\texttt{for i in range(len($E[T']$)):}$'' \label{algline:map_header}
        \State $E[T'_{1, -1}] \gets E[T'] \cdot \text{``}\texttt{[i]}\text{''}$ \label{algline:map_indexing}
        \If{$\mathit{IsStatementNode}(T'_1)$}
            \State $P \gets P \cdot^* \mathit{ToPythonStatement}(T', E, \mathit{map}=1)$ \label{algline:stmt_map}
        \Else
            \State $P \gets P \cdot^* \left(E[T'_{1, -1}] \cdot \text{``}\texttt{=}\text{''} \cdot \mathit{ToPythonExpression}(T', E) \right)$ \label{algline:expr_map}
        \EndIf
    \ElsIf{$\mathit{IsStatementNode}(T')$}
        \State $P \gets P \cdot^* \mathit{ToPythonStatement}(T', E, \mathit{map}=0)$ \label{algline:stmt}
    \EndIf
\EndFor \label{algline:write_loop_end}
\State $P \gets P \cdot \text{``}\texttt{return }\text{''} \cdot E[T]$ \label{algline:write_return}
\State \Return $P$
\EndProcedure
\end{algorithmic}
\end{algorithm}

Algorithm~\ref{alg:translation} describes our procedure for translating a functional program in our DSL to imperative Python code. First, we construct the abstract syntax tree of the DSL program (Line~\ref{algline:make_ast}). 

In our procedure, only the \texttt{empty} and \texttt{if} primitives create new variables, while other primitives are translated to statements or expressions on existing variables. On Line~\ref{algline:assign_vars}, we sort the AST nodes in reverse topological order and assign variable names \texttt{v\{i\}} to \texttt{empty} and \texttt{if} nodes, starting from \texttt{v1}. We also assign names \texttt{a\{i\}} to function parameters, starting from \texttt{a1}. The resulting $V$ is a map from an applicable AST node to its corresponding variable name. 

On Line~\ref{algline:assign_exprs}, we create the mapping $E$ from an AST node to the expression representing the result of the computation at that node. The \texttt{length} and \texttt{index} primitives, integers, and comparison and boolean operators are directly mapped to their corresponding Python expressions. At this stage, we only assign complete expressions and handle partial applications (e.g., the function argument to \texttt{map}) at a later stage. On the other hand, list operations and \texttt{map} will later be translated to Python statements; the key, however, is that the \emph{result} of the computation at that node is just the expression consisting of the list being operated on (e.g., \texttt{v1} in \texttt{v1.append(0)}). Thus, we loop through the AST nodes in reverse topological order, and assign (1) corresponding Python expressions to expression nodes and (2) the expression of the list being operated on to statement nodes.

Next, we start writing the translated Python code $P$. We first write the function header according to the type signature of the program and initialize all variables corresponding to \texttt{empty} nodes at the beginning of the function as a tuple (e.g., \texttt{v1, v2 = [], []}) on Line~\ref{algline:init_vars}. Note that due to Python's scoping rules, variables initialized in conditionals are accessible within the entire function, and do not need to be first declared (and hence why we do not write them on Line~\ref{algline:init_vars}). 

We now loop through the AST nodes in reverse topological order (Lines~\ref{algline:write_loop_start}--\ref{algline:write_loop_end}). We skip any partial nodes, which are those with fewer children than the usual number of parameters (Lines~\ref{algline:partial_skip_start}--\ref{algline:partial_skip_end}); for example, a \texttt{len} node without any children because it is the first child of a \texttt{map} node is partial. For statement nodes (\texttt{append}, \texttt{extend}, \texttt{init}, \texttt{tail}, \texttt{if}) that are not direct children of a \texttt{map} node, we directly translate them to Python according to the rules in Table~\ref{tab:translation_rules} and append the Python statement to $P$ (Line~\ref{algline:stmt}). We write $T'_i$ to mean the $i$\textsuperscript{th} child node of $T'$ (where the index $-1$ refers to the last child); $T'_{i, j}$ refers to the $j$\textsuperscript{th} child node of $T'_i$. We use $\cdot$ to denote string concatenation and $\cdot^*$ to denote string concatenation followed by the concatenation of a newline character (and preceded by an indentation if inside a loop). To translate \texttt{map}, we first write the loop header for the list that is being operated on (Line~\ref{algline:map_header}). Second, we assign the expression of this list (followed by the indexing \texttt{[i]} to retrieve an element of the list) to the expression of the missing argument in the mapping function (Line~\ref{algline:map_indexing}). There are now two cases: one if the mapping function is otherwise translated to statements and the other if the mapping function is \texttt{len} or \texttt{index}. In the first case, we use the same conversion shown in Table~\ref{tab:translation_rules}, except that conditionals are additionally followed by the line \texttt{$E[T'_{1, -1}]$ = $E[T'_1]$} (Line~\ref{algline:stmt_map}). In the second case, we convert the \texttt{len} or \texttt{index} node to a Python expression and prepend it with an assignment to the mapped list element (Line~\ref{algline:expr_map}).

Finally, we return $E[T]$ as the last line of our translated program, which is the expression corresponding to the root node of the AST (Line~\ref{algline:write_return}). 

\begin{table}
    \centering
    \caption{Translation of statement nodes from DSL to Python. $T'$ is the current AST node, and $T'_i$ is the $i$\textsuperscript{th} child node of $T'$. $E$ is the mapping from an AST node to the expression representing the result of the computation at that node.}
    \label{tab:translation_rules}

    \begin{tabular}{ll}
        \toprule
        DSL Program & Python Program \\
        \midrule
        \texttt{append $T'_1$ $T'_2$} & \texttt{$E[T'_2]$.append($E[T'_1]$)} \\
        \texttt{extend $T'_1$ $T'_2$} & \texttt{$E[T'_2]$.extend($E[T'_1]$)} \\
        \texttt{init $T'_1$} & \texttt{$E[T'_1]$.pop()} \\
        \texttt{tail $T'_1$} & \texttt{$E[T'_1]$.pop(0)} \\
        \texttt{if $T'_1$ $T'_2$ $T'_3$} & \makecell[l]{\texttt{if $E[T'_1]$:}\\\texttt{\space\space$E[T']$ = $E[T'_2]$}\\\texttt{else:}\\\texttt{\space\space$E[T']$ = $E[T'_3]$}}\\
        \bottomrule
    \end{tabular}
\end{table}

\subsection{LLM-List} \label{sec:llm_list_details}
We use the following prompts to create the LLM-List dataset. For input generation, we prepend our request with an example to show the model the correct output format. We use the function {\ttfamily \obeyspaces `def add(a, b):{\textbackslash}n    return a + b'} with the description ``returns the sum of two numbers'', and specify the response as \texttt{`3, 5{\textbackslash}n-2, 7{\textbackslash}n0, 0'}. If any generated input contains floating-point numbers or leads to a runtime error, we regenerate the inputs and append the following sentence to the end of the prompt instruction (before the function code): \texttt{`Do NOT include the following inputs:'} followed by a comma-separated list of the erroneous inputs.

\begin{tcolorbox}[title={LLM-List Brainstorming Prompt}]
\begin{lstlisting}
Your task is to brainstorm a list of 100 known / common list functions in Python. These could be standard textbook algorithms or simple utility functions. Some examples are length, reverse, unique, compact, flatten, insert, index, union, tail, permutations, order-by, mean, median, range, argmax.

Each function you come up with must satisfy the following conditions:
- Takes in a list of integers as one of the parameters and returns a list, integer, or boolean after doing some processing on the input.
- Does NOT contain random operations.
- Does NOT involve substantial floating-point operations.
- Does NOT rely on any imports (e.g., numpy or the Python standard library).

Try to have as much variability in the types of operations; for any class or variations of operations, have at most 2-3. Structure your response in the following manner. The name should be a function signature (e.g., length(lst)), and the description should encapsulate the expected behavior of the function.

1. "[name]": "[description]"
2. "[name]": "[description]"
3. "[name]": "[description]"
...
\end{lstlisting}
\end{tcolorbox}

\begin{tcolorbox}[title={LLM-List Code Generation Prompt}]
\begin{lstlisting}
Your task is to write a Python function `@{function_header}@` that @{function_description}@. You may use built-ins, but limit your usage so the function has enough logic in it; you are not allowed to use numpy. Make the logic in your function as explicit as possible, and make sure that the result returned by your function is deterministic. Do not include comments, and do not output any extra information.
\end{lstlisting}
\end{tcolorbox}

\begin{tcolorbox}[title={LLM-List Input Generation Prompt}]
\begin{lstlisting}
You are given a Python function named `@{function_name}@` below, which @{function_description}@. Your goal is to generate 3 simple test inputs for this function that comprehensively test all functionality of the `@{function_name}@` function and produce no errors when executed. Do NOT include any extra information and put each input on a separate line. If the input contains multiple arguments, separate them by commas. Do NOT include floating-point values. Make sure that lists contain only a few elements, but are not empty.

```python
@{function_code}@
```
\end{lstlisting}
\end{tcolorbox}
\clearpage

\subsection{Example Problems} \label{sec:example_programs}
\begin{figure}[h!]
\centering

\begin{tcolorbox}[boxrule=0.5pt, colframe=black, colback=white, width=\textwidth]
\begin{minipage}[t]{0.48\textwidth}
  \centering
  \vspace{-6pt}
  \includegraphics[width=\linewidth]{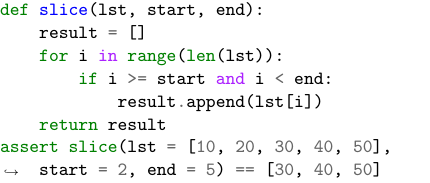}
\end{minipage}
\hfill
\tikz[baseline]{\draw[dotted, line width=0.5pt] (0,0.2) -- (0,-2.97);}
\hfill
\begin{minipage}[t]{0.48\textwidth}
  \centering
  \vspace{-6pt}
  \includegraphics[width=\linewidth]{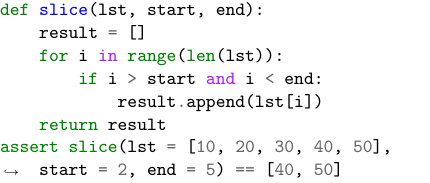}
\end{minipage}

\noindent
\begin{tikzpicture}
    \draw[dotted, line width=0.5pt] (0, 0) -- (\textwidth, 0);
\end{tikzpicture}

\begin{minipage}[t]{0.48\textwidth}
  \centering
  \vspace{-6pt}
  \includegraphics[width=\linewidth]{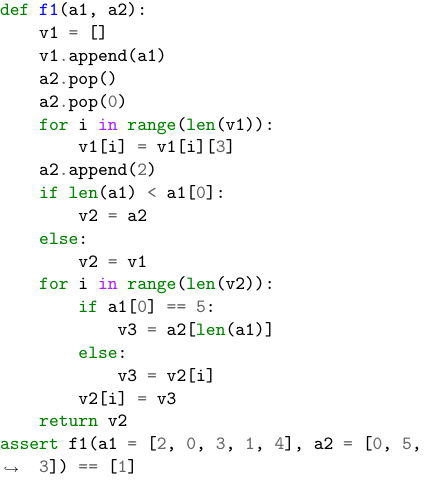}
\end{minipage}
\hfill
\tikz[baseline]{\draw[dotted, line width=0.5pt] (0,0.5) -- (0,-8.12);}
\hfill
\begin{minipage}[t]{0.48\textwidth}
  \centering
  \vspace{-6pt}
  \includegraphics[width=\linewidth]{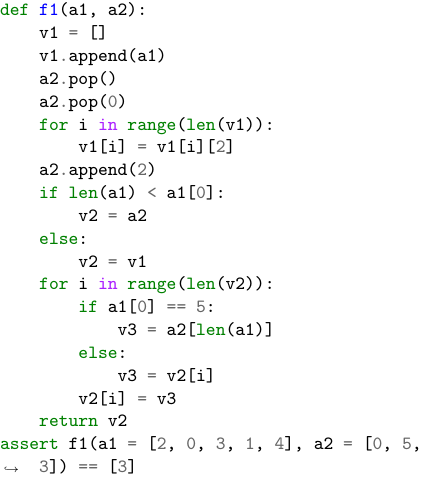}
\end{minipage}

\noindent
\begin{tikzpicture}
    \draw[dotted, line width=0.5pt] (0, 0) -- (\textwidth, 0);
\end{tikzpicture}

\begin{minipage}[t]{0.48\textwidth}
  \centering
  \vspace{-6pt}
  \includegraphics[width=\linewidth]{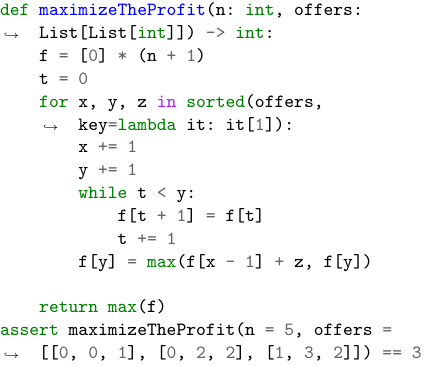}
\end{minipage}
\hfill
\tikz[baseline]{\draw[dotted, line width=0.5pt] (0,0.5) -- (0,-6.1);}
\hfill
\begin{minipage}[t]{0.48\textwidth}
  \centering
  \vspace{-6pt}
  \includegraphics[width=\linewidth]{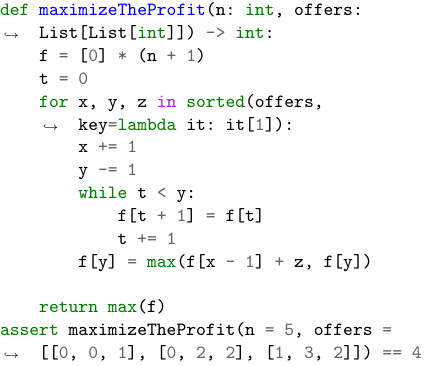}
\end{minipage}
\end{tcolorbox}

\caption{Example original (left column) and mutated (right column) problems from the LLM-List (top row), DSL-List (center row), and LeetCode (bottom row) datasets.}
\label{fig:program_examples_appendix}
\end{figure}

\section{Experimental Details}
\subsection{Benchmarked Models} \label{sec:model_identifiers}
We provide the specific identifiers for the large language models we evaluate in Table~\ref{tab:model_checkpoints}. For each model, we provide its name on HuggingFace (for open-access models) or specific snapshot (for closed-access models).

\begin{table}[h!]
\centering
\caption{Names and identifiers of the large language models used in our evaluation.}
\label{tab:model_checkpoints}

\begin{tabular}{ll}
\toprule
Model Name & Model Identifier \\
\midrule
DeepSeek-Coder-7B & \texttt{deepseek-ai/deepseek-coder-6.7b-instruct} \\
DeepSeek-Coder-33B & \texttt{deepseek-ai/deepseek-coder-33b-instruct} \\
Qwen2.5-Coder-7B & \texttt{Qwen/Qwen2.5-Coder-7B-Instruct} \\
Qwen2.5-Coder-14B & \texttt{Qwen/Qwen2.5-Coder-14B-Instruct} \\
Qwen2.5-Coder-32B & \texttt{Qwen/Qwen2.5-Coder-32B-Instruct} \\
QwQ-32B & \texttt{Qwen/QwQ-32B} \\
DeepSeek-R1 & \texttt{deepseek-ai/DeepSeek-R1} \\
GPT-4o-mini & \texttt{gpt-4o-mini-2024-07-18} \\
GPT-4o & \texttt{gpt-4o-2024-08-06} \\
o3-mini & \texttt{o3-mini-2025-01-31} \\
\bottomrule
\end{tabular}
\end{table}

\subsection{Execution Prediction Prompts} \label{sec:exec_pred_prompts}
We present our execution prediction prompts below. We use the zero-shot prompt for reasoning models and the one-shot prompt for traditional models.

\begin{tcolorbox}[title={Execution Prediction Prompt (Zero-Shot)}]
\begin{lstlisting}
You are given a Python program and an assertion containing an input to a function. Replace the ?? in the assertion with a literal (no unsimplified expressions, no function calls) representing the function's return value for the given input. Execute the program exactly as written, even if it is incorrect or incomplete. For your final answer, provide the full assertion in [ANSWER] and [/ANSWER] tags.

[PYTHON]
@{program}@
assert @{function_name}@(@{input}@) == ??
[/PYTHON]
\end{lstlisting}
\end{tcolorbox}

\begin{tcolorbox}[title={Execution Prediction Prompt (One-Shot)}]
\begin{lstlisting}
You are given a Python program and an assertion containing an input to a function. Replace the ?? in the assertion with a literal (no unsimplified expressions, no function calls) representing the function's return value for the given input. Execute the program exactly as written, even if it is incorrect or incomplete. Execute the program step by step before arriving at an answer, and provide the full assertion with the function output in [ANSWER] and [/ANSWER] tags, following the example.

[PYTHON]
def performOperation(s):
    s = s + s
    return "b" + s + "a"
assert performOperation(s = "hi") == ??
[/PYTHON]
[THOUGHT]
Let's execute the code step by step:

1. The function performOperation is defined, which takes a single argument s.
2. The function is called with the argument "hi", so within the function, s is initially "hi".
3. Inside the function, s is concatenated with itself, so s becomes "hihi".
4. The function then returns a new string that starts with "b", followed by the value of s (which is now "hihi"), and ends with "a".
5. The return value of the function is therefore "bhihia".
[/THOUGHT]
[ANSWER]
assert performOperation(s = "hi") == "bhihia"
[/ANSWER]

[PYTHON]
@{program}@
assert @{function_name}@(@{input}@) == ??
[/PYTHON]
\end{lstlisting}
\end{tcolorbox}

\subsection{Execution Choice Prompts} \label{sec:exec_choice_prompts}
We present our execution choice prompts below. We use the zero-shot prompt for reasoning models and GPT-4o and the one-shot prompt for the other models. We observed that the one-shot prompt prevented GPT-4o from performing chain-of-thought thinking, whereas including an example was beneficial to elicit chain-of-thought thinking from other traditional LLMs.

\begin{tcolorbox}[title={Execution Choice Prompt (Zero-Shot)}]
\begin{lstlisting}
You are given two Python programs below and an assertion containing an input to a function. First, choose either program, whichever one you are more confident in reasoning about. Then, replace the ?? in the assertion with a literal (no unsimplified expressions, no function calls) representing the function's return value for the given input on your chosen program. Execute the program exactly as written, even if it is incorrect or incomplete. For your final answer, output the letter of your chosen program (A or B) and the full assertion in the following json format:
{
    "chosen_program": chosen_program_letter,
    "assertion": full_assertion
}

[PROGRAM_A]\n@{program_a}@\n[/PROGRAM_A]
[PROGRAM_B]\n@{program_b}@\n[/PROGRAM_B]
[ASSERTION]\nassert @{function_name}@(@{input}@) == ??\n[/ASSERTION]
\end{lstlisting}
\end{tcolorbox}

\begin{tcolorbox}[title={Execution Choice Prompt (One-Shot)}]
\begin{lstlisting}
You are given two Python programs below and an assertion containing an input to a function. First, choose either program, whichever one you are more confident in reasoning about. Then, replace the ?? in the assertion with a literal (no unsimplified expressions, no function calls) representing the function's return value for the given input on your chosen program. Execute the program exactly as written, even if it is incorrect or incomplete. Execute the program step by step before arriving at an answer, then output the letter of your chosen program (A or B) and the full assertion in the following json format:
{
    "chosen_program": chosen_program_letter,
    "assertion": full_assertion
}

# Example
[PROGRAM_A]
def performOperation(s):
    first = s[0].upper()
    rest = s[1:].upper()
    return first + rest
[/PROGRAM_A]
[PROGRAM_B]
def performOperation(s):
    first = s[0].upper()
    rest = s[1:].lower()
    return first + rest
[/PROGRAM_B]
[ASSERTION]
assert performOperation(s = 'hELLO') == ??
[/ASSERTION]
[THOUGHT]
First, let's figure out which program I am more confident in reasoning about.

Looking at programs A and B, the difference is in the expression for rest. Program A defines rest as s[1:].upper() while program B defines rest as s[1:].lower(). Program B looks similar to how one might implement the capitalize() function, so I will choose program B as I am more confident in reasoning about this program behavior. 

Now, let's execute the code step by step:

1. The function performOperation is defined, which takes a single argument s.
2. The function is called with the argument 'hELLO', so within the function, s is initially 'hELLO'.
3. The variable first is defined as the upper case of the first character of s, which is 'H'.
4. The variable rest is defined as the lower case of s[1:], which is 'ello'.
4. The function returns first ('H') concatenated with rest ('ello').
5. The return value of the function is therefore 'Hello'.
[/THOUGHT]
{
    "chosen_program": "B",
    "assertion": "assert performOperation(s = 'hELLO') == 'Hello'"
}

# Question
[PROGRAM_A]\n@{program_a}@\n[/PROGRAM_A]
[PROGRAM_B]\n@{program_b}@\n[/PROGRAM_B]
[ASSERTION]\nassert @{function_name}@(@{input}@) == ??\n[/ASSERTION]
\end{lstlisting}
\end{tcolorbox}
\clearpage

\section{Experimental Results} \label{sec:results_appendix}
\begin{figure}[h!]
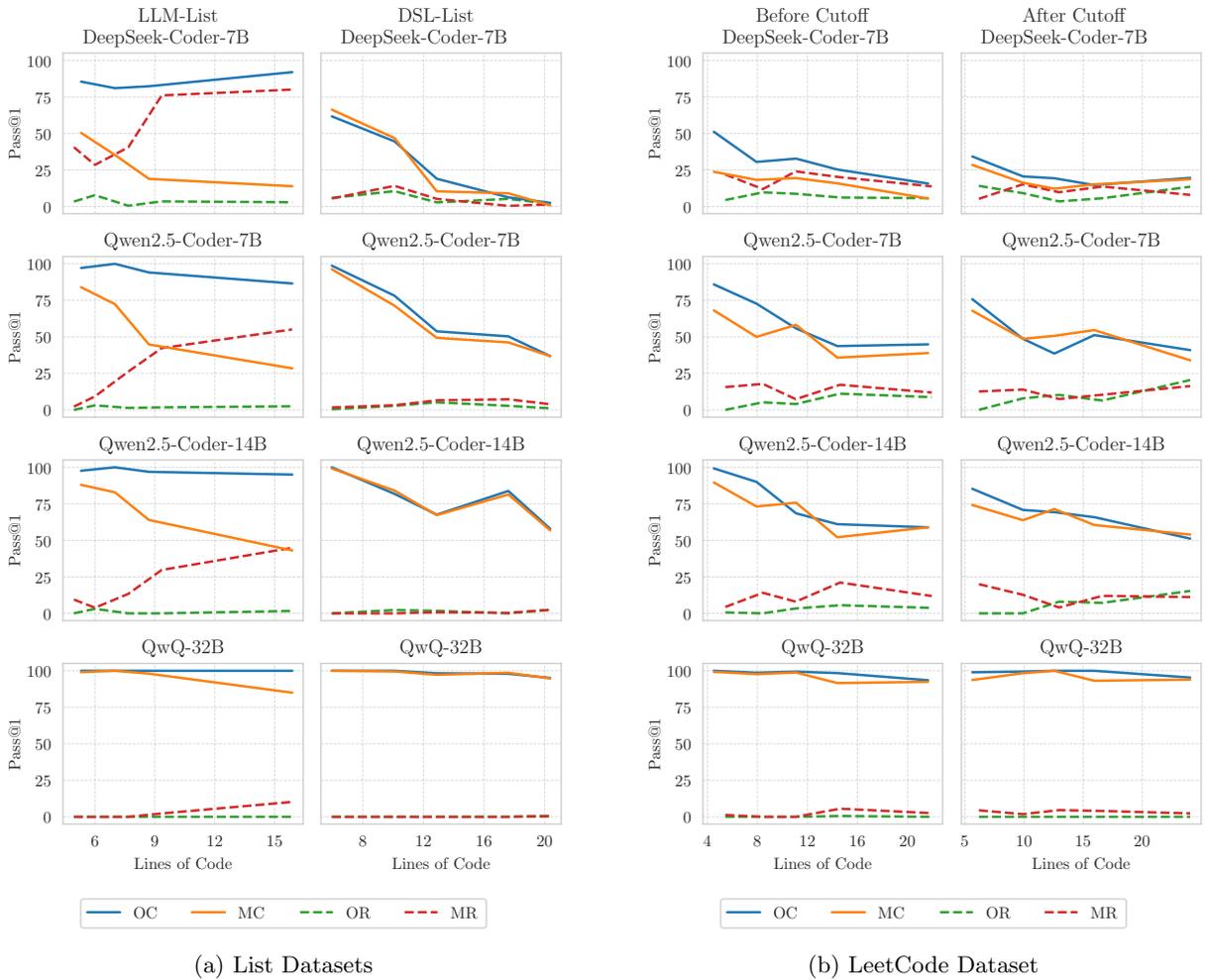

    \centering
    \begin{subfigure}[b]{0.48\textwidth}
        \centering
        \resizebox{\linewidth}{!}{\input{figures/list_open_rest.pgf}}
        \caption{List Datasets}
        \label{fig:list_loc_open_appendix}
    \end{subfigure}
    \hfill
    \begin{subfigure}[b]{0.48\textwidth}
        \centering
        \resizebox{\linewidth}{!}{\input{figures/leetcode_open_rest.pgf}}
        \caption{LeetCode Dataset}
        \label{fig:leetcode_loc_open_appendix}
    \end{subfigure}

    \caption{Execution prediction results on remaining open-access models as a function of lines of code.}
    \label{fig:loc_open_appendix}
\end{figure}

We present the execution prediction results as a function of lines of code for the remaining open-access models not shown in the main text in Fig.~\ref{fig:loc_open_appendix}.

